\documentstyle[epsf,prd,aps]{revtex}
\begin{document}
\draft
\title
{On conformally flat initial data for Einstein equations.}
\author
{   Janusz Karkowski and Edward Malec  }
\address{   Institute of Physics,  Jagiellonian University,
30-059  Cracow, Reymonta 4, Poland.}

\maketitle
\begin{abstract}
We describe     conformally flat initial data, with explicitly given
analytic  extrinsic curvature solving the vacuum momentum constraints. They follow from
a solution of Dain and Friedrich discovered in 2001.
The cylindrically symmetric subcase of the Bowen-York solution is a subclass of this general configuration.

\end{abstract}

The process of solving   initial constraints of Einstein equations
is a difficult task, since they constitute a set of nonlinear  partial differential
equations. Having said that, it is necessary to point out that  these  
equations are undetermined; thus it comes as an embarrassment that so little is known
about their  explicit  solutions in (at least) some simple cases.
The only data that had been available for a long time,
up to our knowledge, are those  of Bowen-York \cite{Bowen} and (discovered much later) Brandt-Seidel \cite{Brandt},
in which vacuum momentum constraints are solved explicitly and it remains to solve only
the Lichnerowicz-York equation.

In the first version of this  short note we presented another set of explicit solutions of vacuum momentum
constraints, obtained by work and guess, that includes a part of the Bowen-York data as a very special subcase.
These appear to be included in a solution found by Dain and Friedrich in 2001 \cite{Dain}. The purpose of this version
is to explain the relation between the two solutions.

Let $\Sigma $ be a Cauchy hypersurface endowed with  an internal metric $g_{ij}$ with
the scalar curvature $R$, and the extrinsic curvature $K_{ij}$. The initial constraint
equations in vacuum read \cite{MTW}
\begin{eqnarray}
R&=&   K_{ij}K^{ij}-\left( K_i^i\right)^2
\nonumber \\
&& \nabla_i\left( K^i_j- g^i_jK^l_l\right) = 0.
\label{1}
\end{eqnarray}
  In the case of maximal slicing condition, $K_i^i=0$, the initial data can be found
relatively easily using the conformal method \cite{York1973}.
 Let $g_{ij}=\Phi^4  \hat g_{ij}$ where  $\hat g_{ij}$ is the Euclidean metric.
The new solution is defined by the following  extrinsic curvature,
 in spherical coordinates,
\begin{eqnarray}
&&K_{23}= K_{13}=0\nonumber
\\
&&K_1^1={1\over r^3\Phi^6}\partial^2_xW \nonumber
\\
&& K_1^2={1\over \sin \theta r^3\Phi^6}\partial_r\partial_xW\nonumber
\\
&&K_2^2={1\over r^2\sin^2\theta \Phi^6}\left[ \partial_r\left( r\partial_rW\right) +{1\over r} \left( x\partial_xW
-W\right) \right] .
\label{3}
\end{eqnarray}
Here $x=\cos \theta $ and $W $ is an arbitrary function of $r$ and $\theta $.
It is easy to check that (\ref{3}) satisfies the momentum constraints, $\nabla_iK^i_j=0$, where
$\nabla_i$ is the covariant derivative in the sense of the metric $g_{ij}$.

This solution generalizes the Bowen-York solution (or, strictly saying, its cylindrically symmetric
part with vanishing angular-momentum) of momentum constraints; the latter corresponds to a  particular choice of
$W$. For instance, a part of the  Bowen-York   extrinsic curvature that is displayed below,
\begin{eqnarray}
K_{ij}&=&{3\over 2r^2 \Phi^2}\left( \hat P_i\hat n_j + \hat P_j\hat n_i -\left( \hat g_{ij}
-\hat n_i\hat n_j\right) \hat P_l\hat n^l \right) +\nonumber\\
&&{3a^2\over 2r^4 \Phi^2}\left( \hat P_i\hat n_j + \hat P_j\hat n_i +\left( \hat g_{ij}
-5\hat n_i\hat n_j\right) \hat P_l\hat n^l \right),
\label{4}
\end{eqnarray}
(with a constant vector ${\vec {\hat P}}$ aligned along the z-axis and $ {\vec {\hat n}}$ being a unit normal
to centered metric spheres in the $\hat g_{ij}$ metric)
can be obtained from (\ref{3}) by substituting
\begin{equation}
W=r {\hat P\over 2}\left( x^3-3x\right) \left(1 -{a^2\over r^2}\right) .
\label{5}
\end{equation}
Another special choice of the function $W$,
\begin{equation}
W=  {\hat P \over 2}\left( x^2+1\right) ,
\label{6}
\end{equation}
recovers the well known  spherically symmetric solution of the momentum constraint
\begin{eqnarray}
&&K^1_1={\hat P\over r^3\Phi^6}=-2K_2^2=-2K^3_3.
\nonumber\\
&&K_1^2=K_1^3=K^2_3=0.
\label{7}
\end{eqnarray}
We would like to point out  that (\ref{3}) is a cylindrically symmetric subset of a result
obtained by Dain and Friedrich \cite{Dain}, who found (using the Newman-Penrose edth symbols)
the general extrinsic curvature corresponding to the conformally flat initial data.
Their  results imply ours, if one assumes (in the notation of \cite{Dain}, Sec. 4)
\begin{equation}
W=2\lambda_2^R(r, \theta )\sin^2\theta
\label{7a}
\end{equation}
and equates to zero all functions other than $\lambda_2^R$ in their equations (80 - 82).

There exists yet other set of explicit solutions, in which
\begin{eqnarray}
K_{13}&=&{1\over r^2}\partial_x Z(r,x) \nonumber\\
K_{23}&=&{1\over \sin \theta }\partial_r Z(r,x)
\label{8}
\end{eqnarray}
and the remaining components of $K_{ij}$ do vanish. The function $Z$   does not depend on the
$\phi -$angle variable, but otherwise is arbitrary. That is the conformally flat
version of the Brandt-Seidel solution reported in \cite{Brandt}; its conformally flat generalization
can be also found in \cite{Dain}. A particular form of
$Z$ generates the angular-momentum part of the Bowen-York solution. Any linear combination
of the two solutions,   (\ref{3}) and (\ref{8}), obviously solves the momentum constraints.

The remaining hamiltonian constraint reduces to
the Lichnerowicz-York equation $\Delta \Phi = -{1\over 8} K_{ij}K^{ij}\phi^5$ (here $\Delta $
 is the flat laplacian), which determines
the conformal factor $\Phi $.
One could simulate  "multi-black hole" initial data  by the appropriate choice of $W$.
The global energy-momentum can be found from standard formulae
\begin{eqnarray}
E&=&{-1\over 2\pi } \int_{S_{\infty }}d^2S^i\nabla_i\phi ,
\nonumber \\
P_i&=&{1\over 8\pi } \int_{S_{\infty }}d^2S^iK_{ij}.
\label{2}
\end{eqnarray}
The asymptotic mass reads $m=\sqrt{E^2-P_iP^i}$.
It is clear that there exists a large set of $W$'s that generate isolated systems
with finite global energy and momentum.

\begin{acknowledgements}
This work was partly supported by  the KBN grant 2 PO3B 00623.
The authors thank Sergio  Dain and Helmut Friedrich for kind information about their results.
\end{acknowledgements}

\end{document}